\documentclass[10pt]{article}
\usepackage{graphicx}
\def\Title#1{\begin{center} {\Large #1 } \end{center}}
\def\Author#1{\begin{center}{ \sc #1} \end{center}}
\def\Address#1{\begin{center}{ \it #1} \end{center}}

\newcommand\pubblock{\rightline{\begin{tabular}{l} Proceedings of the Fifth Annual LHCP\\
         \pubdate  \end{tabular}}}

\newenvironment{Abstract}{\begin{quotation} \begin{center} 
             \large ABSTRACT \end{center}\bigskip 
      \begin{center}\begin{large}}{\end{large}\end{center} \end{quotation}}

\newenvironment{Presented}{\begin{quotation} \begin{center} 
             PRESENTED AT\end{center}\bigskip 
      \begin{center}\begin{large}}{\end{large}\end{center} \end{quotation}}

\def\beq{\begin{equation}}
\def\eeq#1{\label{#1}\end{equation}}
\def\eeqn{\end{equation}}

\def\beqa{\begin{eqnarray}}
\def\eeqa#1{\label{#1}\end{eqnarray}}
\def\eeqan{\end{eqnarray}}

\let\bar=\overbar

\def\Dslash{\not{\hbox{\kern-4pt $D$}}}
\def\dslash{\not{\hbox{\kern-2pt $\del$}}}

\def\msb{{\bar{\ssstyle M \kern -1pt S}}}

\usepackage{color}
\usepackage{subfig}

\newcommand\pubdate{\today}

\def\affiliation{
on behalf of the International Particle Physics Outreach Group, \\
Kirchhoff-Institute for Physics \\
Heidelberg University, Germany}

\begin{document}
% \linenumbers

% large size for the first page
\large
\begin{titlepage}
\pubblock

%% Change the title, name, abstract
%% Title 
\vfill
\Title{Particle Physics Masterclasses for the International Day of Women and Girls in Science}
\vfill

%  if you need to add the support use this, fill the \support definition above. 
%   \Author{ FIRSTNAME LASTNAME \support }
\Author{ Julia Isabell Djuvsland }
\Address{\affiliation}
\vfill
\begin{Abstract}
On the occasion of the UN International Day of Women and Girls in Science (February 11) Masterclass activities were 
launched by the International Particle Physics Outreach Group to support and promote the access of women and girls to science 
education and research activities. Universities and research laboratories organised 10 Masterclasses for girls on February 10 and 
11, with participation from Barcelona, Cagliari, Cosenza, Heidelberg, Madrid, Paris, Prague, Rio de Janeiro, and Sao Paulo. About 
300 girls participated in the events and analysed LHC data while being tutored by female scientists. Three video 
conferences with CERN were held where the girls could talk to CERN women scientists and learn about the careers of these role 
models.
\end{Abstract}
\vfill

% DO NOT CHANGE 
\begin{Presented}
The Fifth Annual Conference\\
 on Large Hadron Collider Physics \\
Shanghai Jiao Tong University, Shanghai, China\\ 
May 15-20, 2017
\end{Presented}
\vfill
\end{titlepage}
\def\thefootnote{\fnsymbol{footnote}}
\setcounter{footnote}{0}
%

% normal size for the rest
\normalsize 

%% Your paper should be entered below. 

\section{Introduction}
Gender equality has come a long way and women enjoy the same legal rights as men in many countries today. 
Yet, there are still far fewer female than male students and researchers, while men and women make up equal parts of our 
society. 
To counteract this fact, the United Nations General Assembly declared 11 February as the International Day of Women and Girls in 
Science (IDWGS). 
Awareness-raising events are encouraged on this date and this year the International Particle Physics Outreach Group (IPPOG) 
followed this call by organising a special edition of their International Masterclasses. 
IPPOG is an international network of scientists, science educators and communication specialists with the goal of conveying 
particle physics to the general public and to improve science education. 
The International Masterclasses are a one-day outreach event and established tool to communicate particle physics to high school 
students. 
To do justice to the UN's call, the IDWGS edition of the International Masterclasses~\cite{IDWGSmasterclasses} was targeted at 
female pupils only.
They were mainly organised and tutored by female scientists in order to provide role models inspiring the participating girls 
to enter university and to take up a career in science. 

\section{The International Masterclasses}

The IPPOG Masterclasses were developed to convey particle physics to high-school students aged between 15 and 19 years. 
They are hosted each spring by about 216 different universities and laboratories located in 52 countries across the globe. 
The local researchers organise the events together with IPPOG and give yearly about 13.000 participants the possibility to 
be particle physicists for one day. 
A typical masterclass is composed of an introduction to particle physics, a hands-on part and a video conference with 
international participants. 
Yet, the exact planning of the day is up to the organising researchers.
Typically, the introduction to particle physics is performed through organising one or two lectures and sometimes also 
includes lab visits. 
After lunch the hands-on part takes place, where the pupils analyse real physics data recorded by one of the LHC detectors. 
The results of the analyses are discussed with the pupils and finally presented to other participants of the Masterclasses 
programme during the video conference. The latter is hosted either by CERN or one of the North American laboratories, Fermilab 
or TRIUMF.
This conveys the international spirit of particle physics and is usually very popular with the students. 

\subsection{Brief History}

The idea of this one-day outreach event first came up in the year 1996 in a discussion between Ken Long and Roger 
Barlow~\cite{Courier0114}.
One year later the first Masterclasses were held at 7 institutes in the United Kingdom. 
They were using data from the OPAL and DELPHI experiments at LEP for the analysis during the hands-on part. 
The programme expanded steadily from there on and in 2005 it was 
adopted for all of Europe by IPPOG's predecessor EPPOG. In 2006 the USA joined with their QuarkNet program and since 2011 the 
International Masterclasses are completely based on the data of the LHC experiments~\cite{Courier0514}. Now, in the spring of 
2017, the International Masterclasses were organised for the first time exclusively for girls following the UN's call for 
contributions to IDWGS.

\subsection{IDWGS Masterclasses}
The International Masterclasses have proven to be a powerful tool in sparking interest for research in young people and in 
encouraging them to pick up a university education. This is in line with what the UN aims at inspiring in girls through its 
establishment of IDWGS and thus the organisation of a special edition of the International Masterclasses was decided by IPPOG on 
the occasion of this day in 2017. In order to involve the female students in the best possible way, IPPOG encouraged the 
organisation of the IDWGS Masterclasses exclusively for girls and, as far as possible, under the supervision of female 
scientists. Ten institutes from Europe and South America followed this call and offered about 320 pupils the chance to 
participate. 
At CERN, 3 video conferences were organised (two on February 10 and one on February 11) and led by female scientists.
This way, the girls met several potential role models during the day.
The 10 institutes that participated in this first edition of the IDWGS Masterclasses were: Barcelona, Cagliari, Cosenza, 
Heidelberg, Madrid, Paris, Prague, Rio de Janeiro, and Sao Paulo. All institutes performed the standard programme foreseen for the 
International Masterclasses and some extended the programme by additional talks or events. 
This is highlighted in the following, where more details on the events of some of the institutions are given.

\begin{figure}[t]
\centering
 \centering
 \subfloat{\includegraphics[width=.58\textwidth]{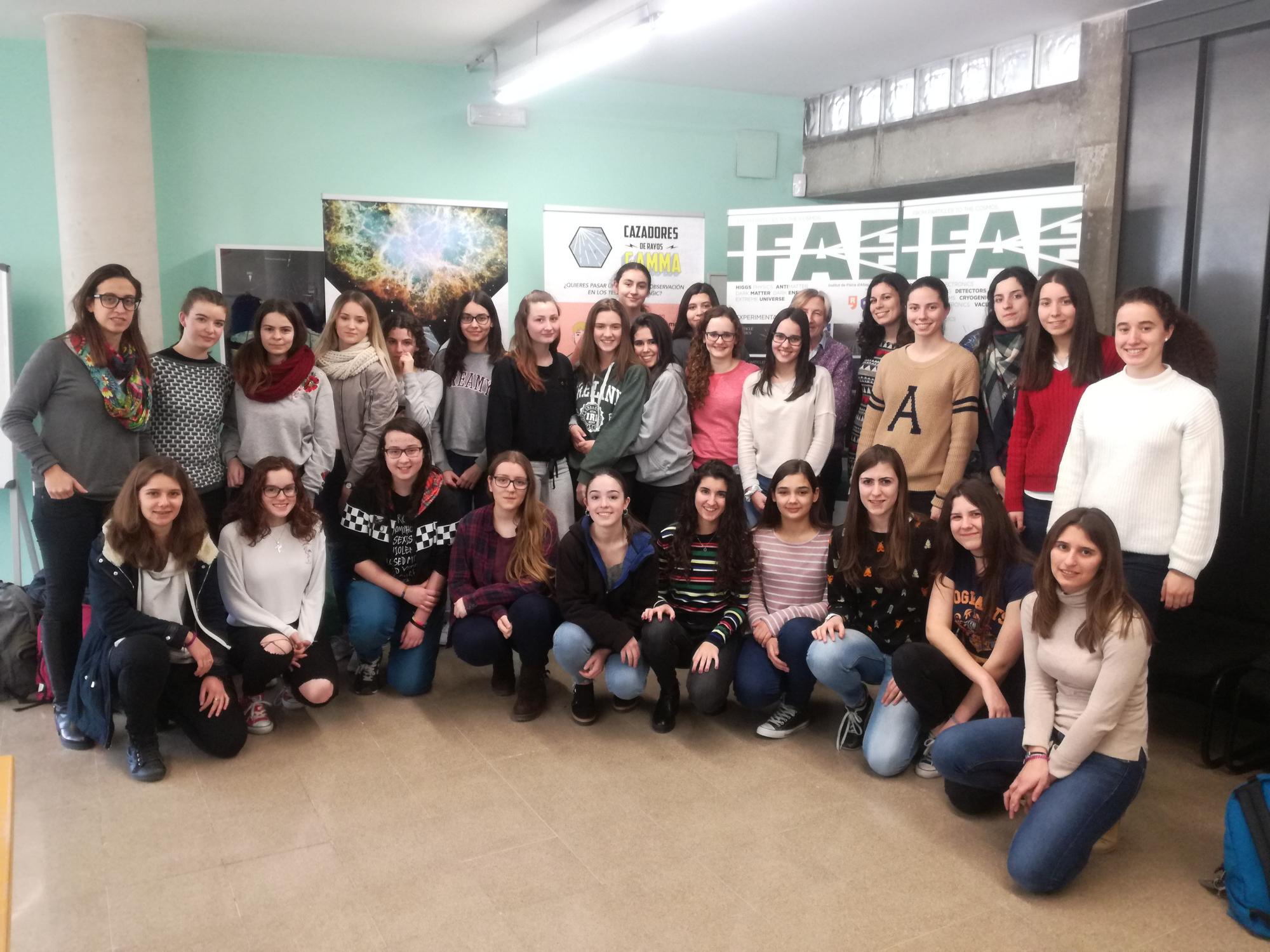}\label{fig:barcaGirls}} \qquad
 \subfloat{\includegraphics[width=.31\textwidth]{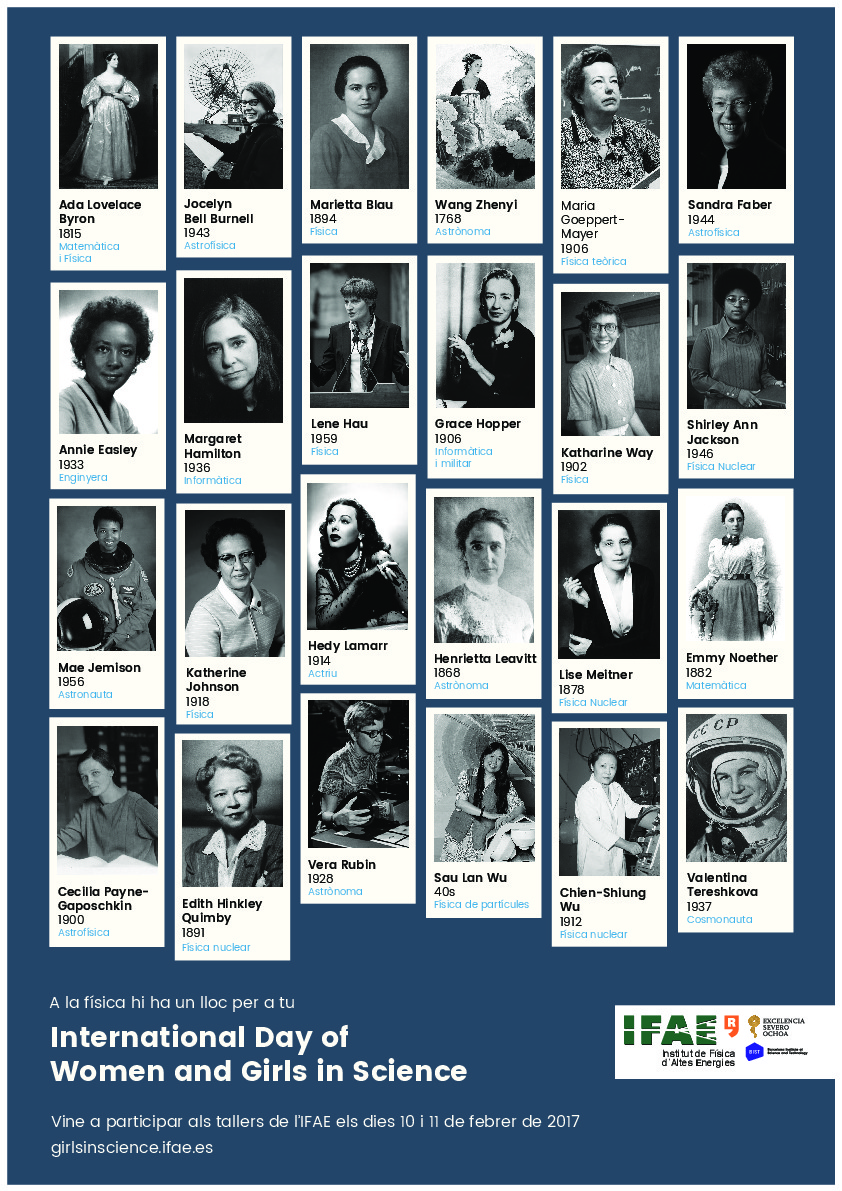}\label{fig:triplFeynm}}
 \caption{Participants of the Masterclasses in Barcelona and poster announcing the event.}
  \label{fig:barcelona}
\end{figure}

\bigskip
\textbf{Institut de Fisica d'Altes Energies, Barcelona}\\
%  (IFAE),  The Barcelona Institute of Science and Technology (BIST)
In Barcelona, a 2-day event~\cite{barcaWebpage} was organised with the International Masterclasses on February 10 and 
astroparticle physics masterclasses on February 11. 
The former included a visit to the local Tier-1 data centre PIC and the latter was performed with the project \textit{Gamma Ray 
Hunters}~\cite{gammaRayHunters}. 
All tutors were women and around 40 girls participated.
A group picture of them is shown in Figure~\ref{fig:barcelona} along with a poster promoting the event.

%%%%%%%%%%%%%%%%%%%%%%%%%%%%%%%%%%%%%%%%%%%%%%%%%%%%%%%%%%%%%

\begin{figure}[t]
\centering
 \centering
 \subfloat[]{\includegraphics[width=.48\textwidth]{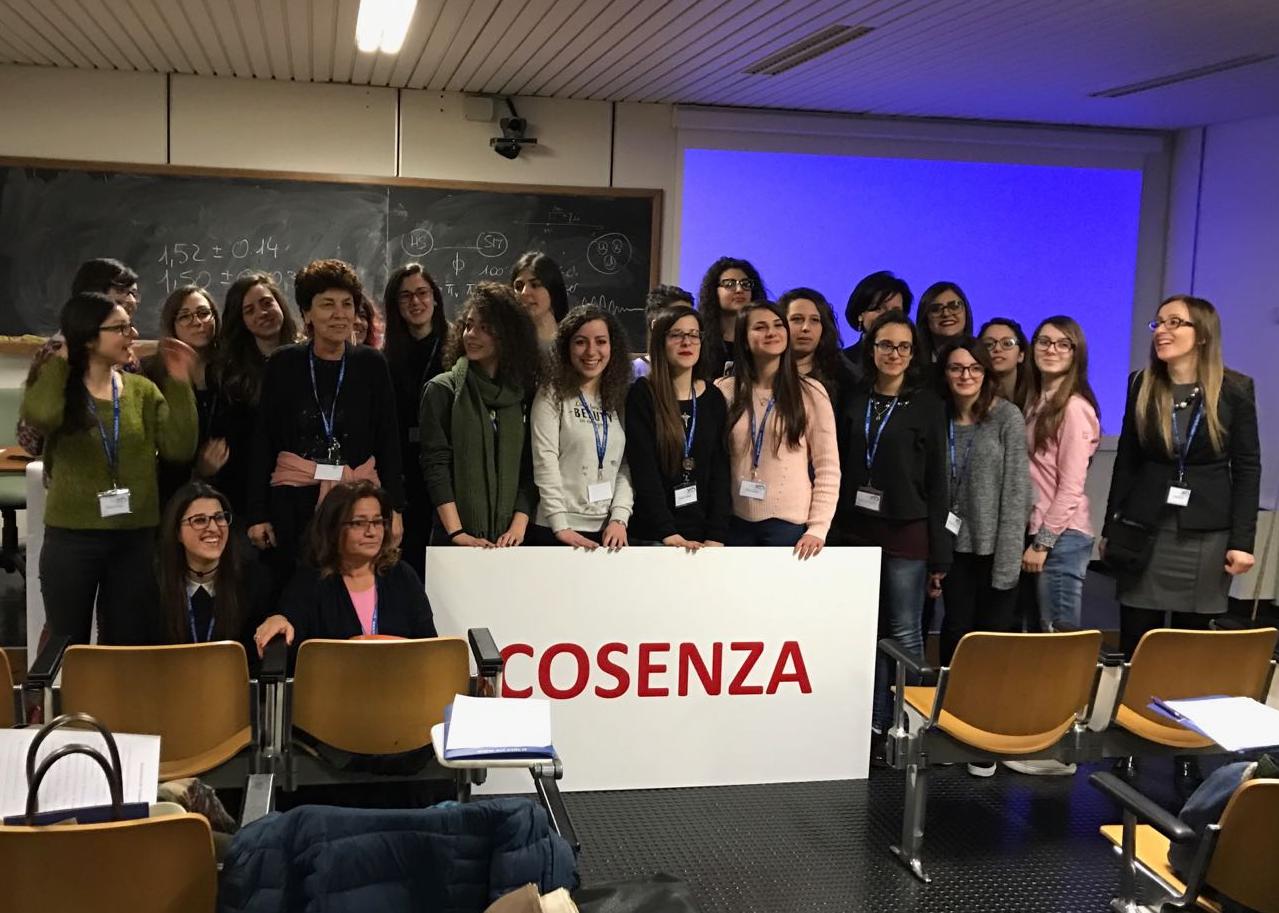}\label{fig:cosenzaGirls}} \quad
 \subfloat[]{\includegraphics[width=.48\textwidth]{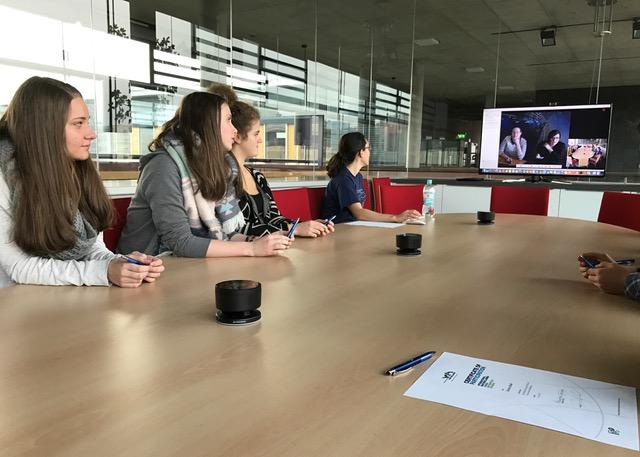}\label{fig:videoHD}}
 \caption{Group picture of the participants in Cosenza~\protect\subref{fig:cosenzaGirls}~\cite{cosenzaWebpage} and impression of 
the video conference held in Heidelberg \protect\subref{fig:videoHD}.}
  \label{fig:cosenBerg}
\end{figure}

\bigskip
% \textbf{Cagliari}
\textbf{Dipartimento di Fisica, Universit\`{a} della Calabria, Cosenza}\\
In Cosenza pupils from various parts of the Calabria region participated in their edition of the IDWGS 
Masterclasses~\cite{cosenzaWebpage}.
Figure~\ref{fig:cosenzaGirls} shows a group picture of the participants and organisers. 
The day started off with a presentation about the situation of women in science and was then followed by the standard 
Masterclasses programme with only female supervisors.
While the organisers also wanted to involve university students at the beginning of their academic studies to encourage them, 
they found this hard to accomplish, as the beginning of February is the period in which the exams are taken. After realising 
this, the organisers requested the acknowledgement of 11 February as a special day of the physics department dedicated to girls, 
in agreement with the UN resolution. This request was acknowledged by the Council of Physics such that this day will receive 
special attention also in the upcoming years.
%  Due to the short time we invited only three schools (10 students each) plus joung univ girls.

%%%%%%%%%%%%%%%%%%%%%%%%%%%%%%%%%%%%%%%%%%%%%%%%%%%%%%%%%%%%%

\bigskip
\textbf{Kirchhoff-Institut f\"ur Physik, Heidelberg}\\
The Heidelberg edition of the IDWGS Masterclasses~\cite{heidelbergWebpage} was also well received by the participants. 
The standard Masterclasses programme was complemented by an informal lunch with two female scientists where the 
girls could ask about university studies and careers in science. A snapshot from the video conference held that day can be seen 
in Figure~\ref{fig:videoHD}.

\bigskip
\textbf{Centro de Investigaciones Energéticas, Medioambientales y Tecnol\'{o}gicas, Madrid}\\
In Madrid~\cite{madridWebpage} the standard Masterclasses programme was extended by a discussion session during which female 
scientists from the department talked about their careers and research activities. In total 40 pupils participated and were 
supervised by female researchers.

\begin{figure}[b]
\centering
 \centering
 \subfloat{\includegraphics[width=.375\textwidth]{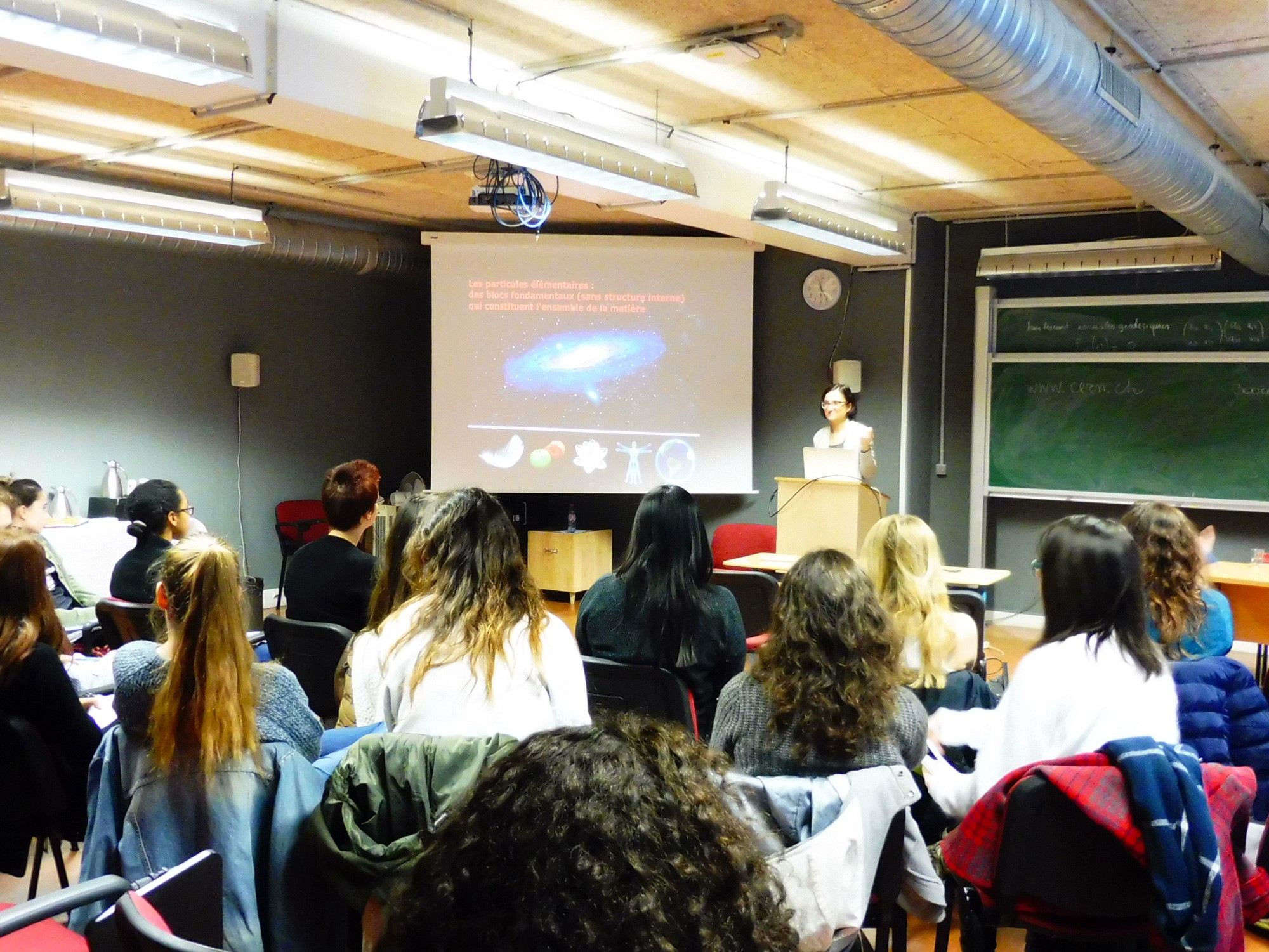}\label{fig:paris1}} \quad
 \subfloat{\includegraphics[width=.375\textwidth]{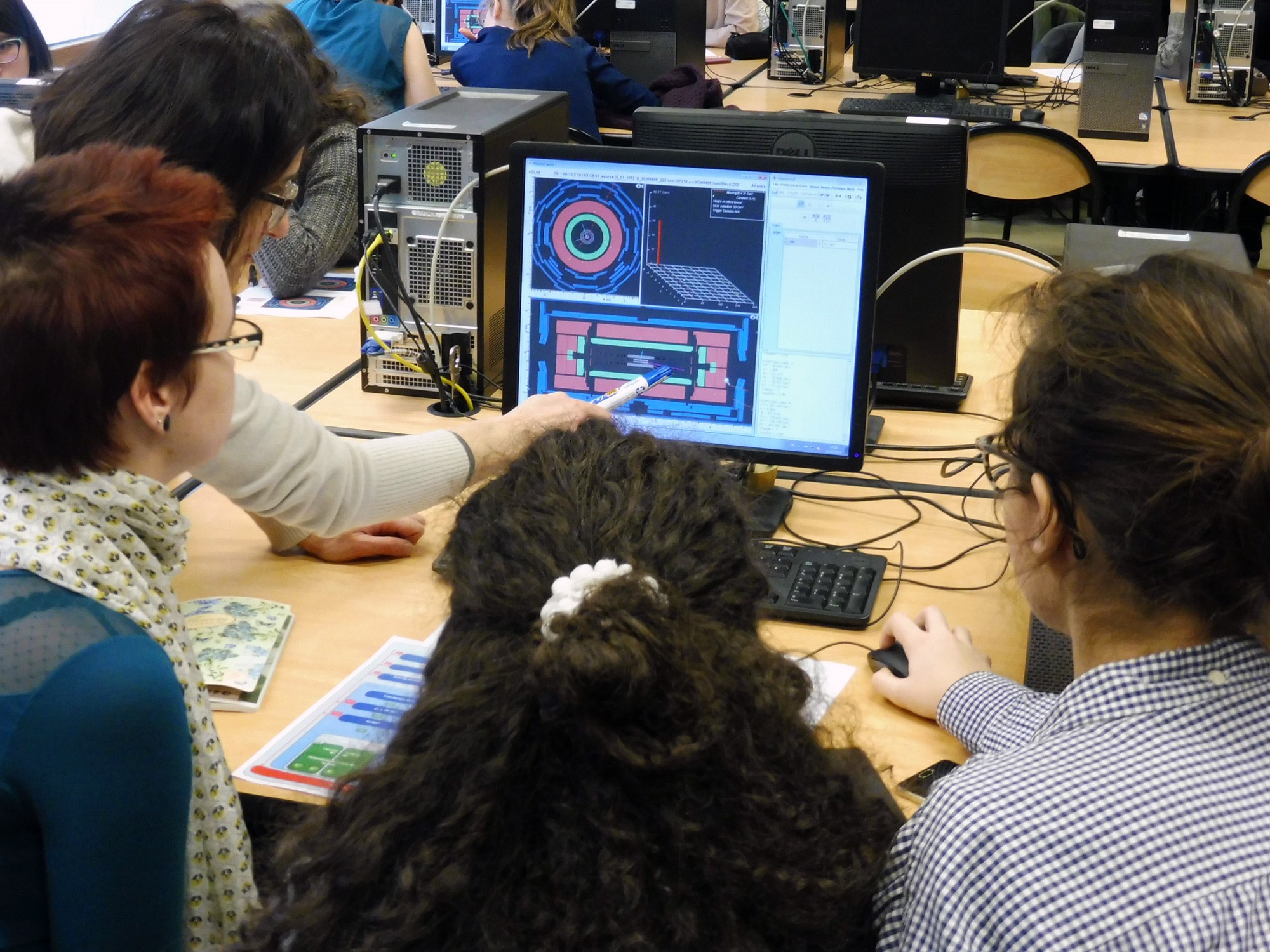}\label{fig:paris2}} \quad
 \subfloat{\includegraphics[width=.2\textwidth]{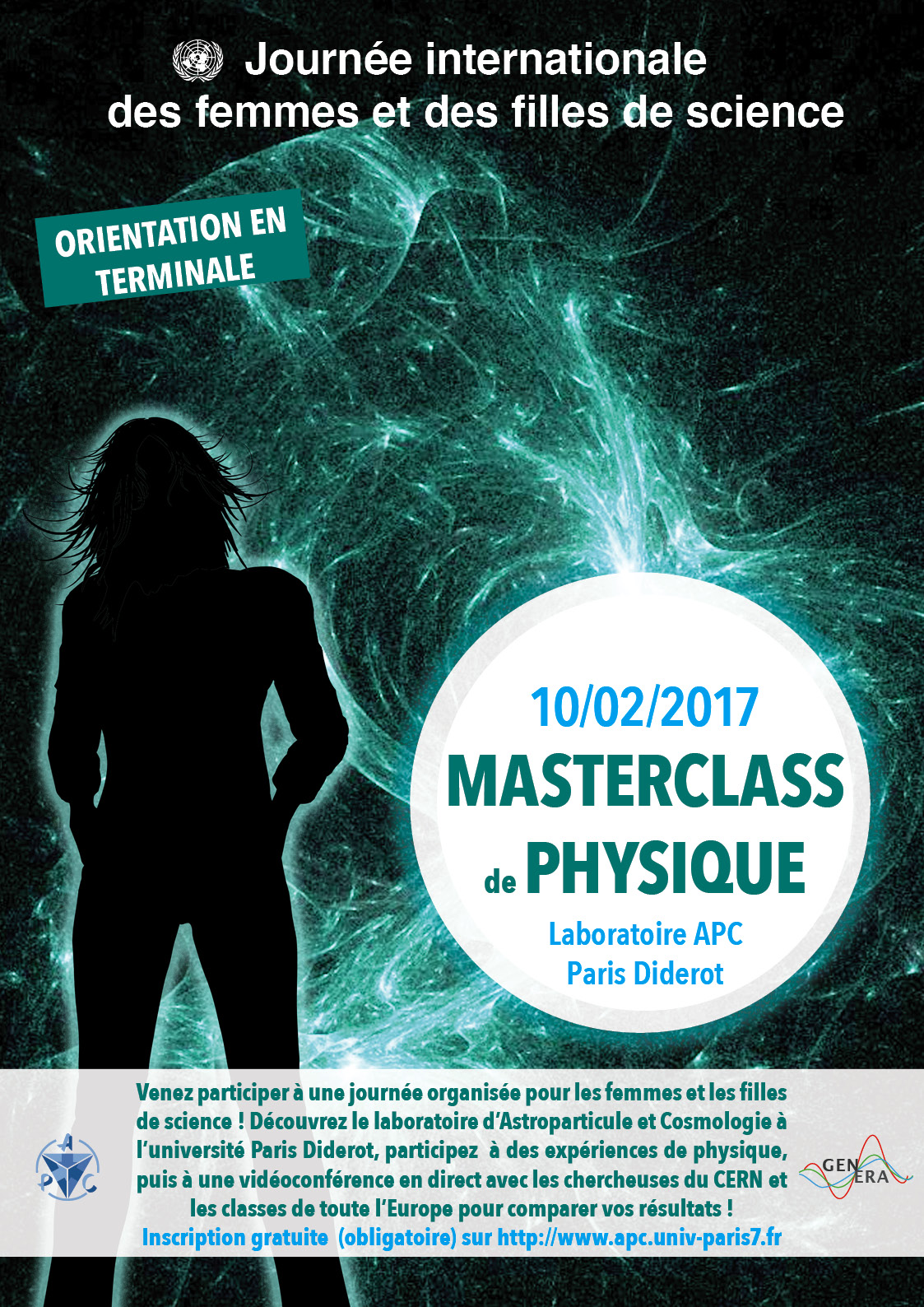}\label{fig:paris3}}
 \caption{Pictures from the IDWGS Masterclasses held in Paris along with the poster announcing the event.}
  \label{fig:paris}
\end{figure}

\bigskip
\textbf{Laboratoire Astroparticule et Cosmologie, Paris}\\
Apart from the standard Masterclasses programme, in Paris two young female researchers (a PhD student and an engineer) 
introduced themselves to the pupils, talked about their professional path and joined the 24 participants for lunch for informal 
discussions~\cite{parisWebpage}. Some impressions of the event are given in Figure~\ref{fig:paris}
%%%%%%%%%%%%%%%%%%%%%%%%%%%%%%%%%%%%%%%%%%%%%%%%%%%%%%%%%%%%%

\begin{figure}[htb]
\centering
 \centering
 \subfloat[]{\includegraphics[width=.55\textwidth]{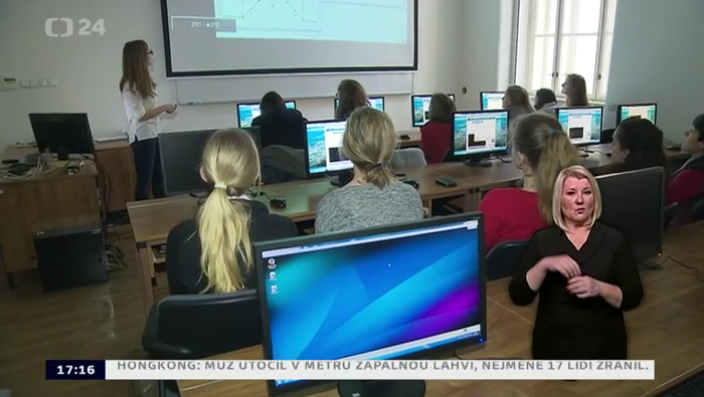}\label{fig:pragueTV}} \quad
 \subfloat[]{\includegraphics[width=.35\textwidth]{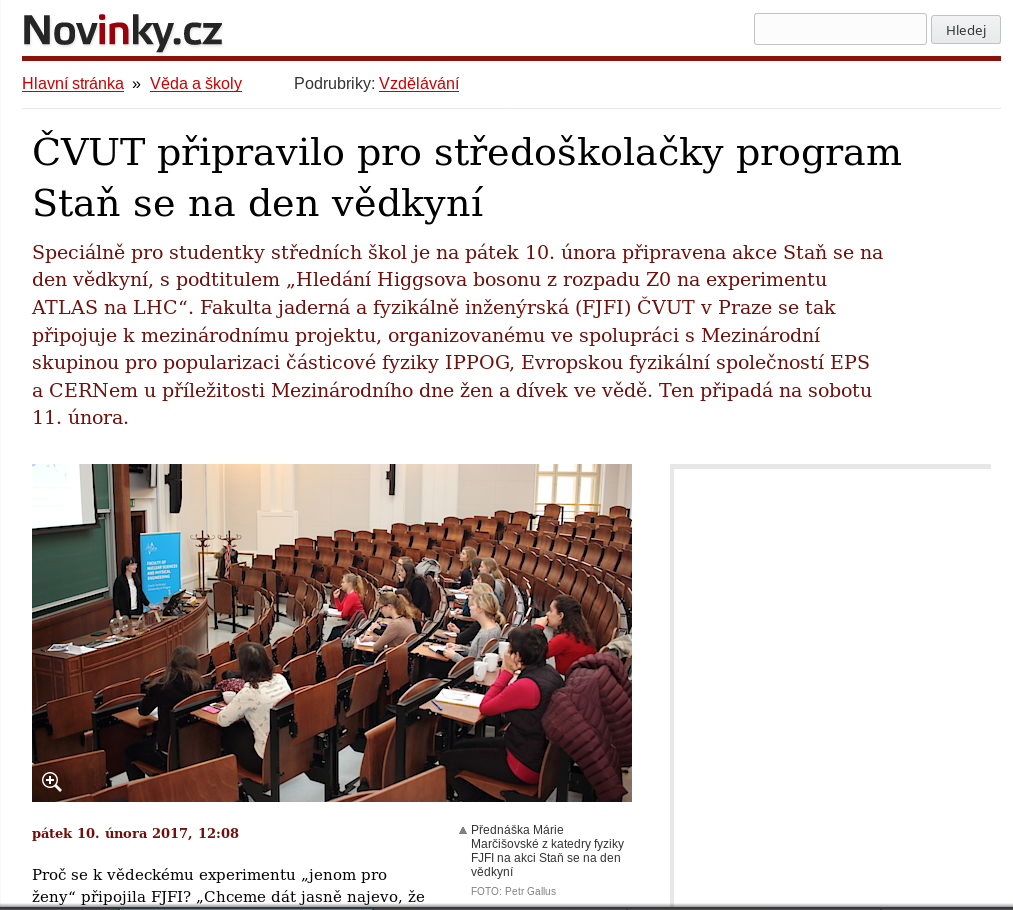}\label{fig:pragueNews}}
 \caption{Screenshot of the TV clip~\cite{pragueTv} reporting about the IDWGS Masterclasses in Prague 
\protect\subref{fig:pragueTV} and from a newspaper article~\cite{prague2} describing the event~\protect\subref{fig:pragueNews}.}
  \label{fig:prague}
\end{figure}

\bigskip
\textbf{Czech Technical University, Prague}\\
In Prague, 19 High School and young University students joined for the International Masterclasses on IDWGS. The event attracted 
quite some media coverage with 2 articles in Czech newspapers\cite{prague1,prague2} and a reporting team of the Czech public 
television being present all day. They continuously reported about the event online and compiled a several minutes 
long TV clip shown on the news~\cite{pragueTv}. 
Figure~\ref{fig:pragueTV} shows a snapshot of the report and Figure~\ref{fig:pragueNews} shows the beginning of one of the 
newspaper articles.

%%%%%%%%%%%%%%%%%%%%%%%%%%%%%%%%%%%%%%%%%%%%%%%%%%%%%%%%%%%%

\bigskip

\textbf{Instituto de Fisica, Rio de Janeiro}\\
In Rio de Janeiro, two IDWGS Masterclasses were organised: one by the physics institute of Universidade do Estado do Rio de 
Janeiro and one by the physics institute of Universidade Federal do Rio de Janeiro. Although they were challenged by the 
political situation in their country, they offered the pupils the opportunity to get to know women scientists and to get in touch 
with current day research.

% \textbf{Sao Paulo}

\section{Conclusions}
This pilot project of organising International Masterclasses on IDWGS especially for girls was a big success. The organisers 
received positive feedback and enjoyed tutoring the girls. During the regular International Masterclasses being held each 
year, the typical number of participants is around 60. Due to the gender preselection and the recent establishment of the IDWGS 
Masterclasses, the groups were smaller on average. This was found to be an advantage, as it allowed for more 
intense discussions with the individual participants. All in all, the IDWGS were so well received that the programme will 
be continued in the upcoming years.
Anyone interested in participating can contact Uta Bilow, the main organiser of this event, via email 
(uta.bilow@tu-dresden.de).

\section{Summary and Outlook}
A special edition of the International Masterclasses was organised by IPPOG on occasion of the UN International Day of Women and 
Girls in Science.
They were targeted at female pupils and mostly carried out by women in order to provide role models for the girls. 
The events were well received and so the efforts will be continued in the upcoming years to inspire even more girls to take up 
university studies and a career in research.

\end{document}